\documentclass[12pt,preprint]{aastex}

\usepackage{psfig}

\newcommand\mzon   {M$_{\odot}$}
\newcommand\pp     {$\pm$}

\newcommand\Lunit   {ergs s$^{-1}$}
\newcommand\funit   {ergs cm$^{-2}$ s$^{-1}$}

\begin{document}

\title{The hard quiescent spectrum of the neutron-star
X-ray transient EXO 1745--248 in the globular cluster Terzan 5}

\author{Rudy Wijnands\altaffilmark{1,2}, Craig O. Heinke\altaffilmark{3}, 
David Pooley\altaffilmark{4}, Peter D. Edmonds\altaffilmark{3}, Walter
H. G. Lewin\altaffilmark{4}, Jonathan E. Grindlay\altaffilmark{3},
Peter G. Jonker\altaffilmark{6,3}, Jon M. Miller\altaffilmark{3}}

\altaffiltext{1}{School of Physics and Astronomy, 
University of St Andrews, North Haugh, St Andrews, Fife, KY16 9SS,
Scotland, UK}

\altaffiltext{2}{
Present address: Astronomical Institute ``Anton Pannekoek'',
University of Amsterdam, Kruislaan 403, 1098 SJ, Amsterdam, The
Netherlands; rudy@science.uva.nl}

\altaffiltext{3}{Harvard-Smithsonian Center for Astrophysics, 
60 Garden Street, Cambridge, MA 02139, USA}

\altaffiltext{4}{Center for Space Research, Massachusetts Institute of
Technology, 77 Massachusetts Avenue, Cambridge, MA 02139, USA}

\altaffiltext{5}{Institute of Astronomy, Madingley Road, CB3 0HA, Cambridge, 
UK}

\begin{abstract}

We present a {\itshape Chandra} observation of the globular cluster
Terzan 5 during times when the neutron-star X-ray transient EXO
1745--248 located in this cluster was in its quiescent state. We
detected the quiescent system with a (0.5--10 keV) luminosity of $\sim
2 \times 10^{33}$ \Lunit. This is similar to several other
neutron-star transients observed in their quiescent states. However,
the quiescent X-ray spectrum of EXO 1745--248 was dominated by a hard
power-law component instead of the soft component that usually
dominates the quiescent emission of other neutron-star X-ray
transients. This soft component could not conclusively be detected in
EXO 1745--248 and we conclude that it contributed at most 10\% of the
quiescent flux in the energy range 0.5--10 keV. EXO 1745--248 is only
the second known neutron-star transient whose quiescent spectrum is
dominated by the hard component (SAX J1808.4--3658 is the other
one). We discuss possible explanations for this unusual behavior of
EXO 1745--248, its relationship to other quiescent neutron-star
systems, and the impact of our results on understanding quiescent
X-ray binaries. We also discuss the implications of our results on the
way the low-luminosity X-ray sources in globular clusters are
classified.
\end{abstract}


\section{Introduction \label{section:introduction}}

Low-mass X-ray binaries (LMXBs) are binary systems in which a compact
object (either a neutron star or a black hole) accretes matter from a
low-mass companion star. The neutron-star X-ray transients are a
sub-group of LMXBs which spend most of their time in a quiescent state
during which hardly any or no accretion onto the neutron star takes
place. Occasionally, huge increases in the mass-accretion rates occur
making those systems very luminous (with X-ray luminosities up to
$\sim 10^{38}$ \Lunit). In their quiescent states, neutron-star X-ray
transients can still be detected with sensitive X-ray instruments at
observed luminosities (0.5--10 keV) of $\sim10^{32}$~\Lunit~to a few
times $10^{33}$ \Lunit~(see, e.g., van Paradijs et al. 1987; Asai et
al. 1996, 1998). Usually, the spectra of those systems in their
quiescent states are dominated by a soft component below 1 keV which
is thought to be thermal emission from the neutron-star surface due to
the cooling of the neutron-star core which has been heated during the
outbursts (Campana et al. 1998; Brown, Bildsten \& Rutledge 1998). An
additional power-law shaped component which dominates above a few keV
might be present in the quiescent spectra. This hard component can
contribute up to half of the emission in the 0.5--10 keV energy range
(e.g., Asai et al. 1998; Rutledge et al. 2001a). However, this
power-law component cannot be detected in all observed quiescent
neutron-star systems and the flux ratio of the power-law component to
the thermal component varies considerably between sources.  The origin
of this hard component is not well understood but it might be due to
residual accretion down to the magnetospheric radius or it may be a
result of the pulsar emission mechanism being active (e.g., Stella et
al. 1994; Campana et al. 1998; Campana \& Stella 2000; Menou \&
McClintock 2001).

Currently, only one neutron-star X-ray transient does not follow this
typical quiescence behavior. Campana et al.~(2002) found that the
accretion-driven millisecond X-ray pulsar SAX J1808.4--3658 had a very
low quiescent luminosity of only $\sim5 \times 10^{31}$ \Lunit~and its
quiescent spectrum was dominated by a hard component with no
significant detection of the thermal component. They suggested that
the very low luminosity of the thermal component might indicate that
the neutron star in SAX J1808.4--3658 is rather cool. Such a cold
neutron star suggests that enhanced cooling processes are active in
its core, possibly because it is a relatively massive neutron star
(with a mass of $>1.7$ \mzon; Colpi et al. 2001). They also suggested
that the hard spectral component possibly originates in the shock
between the wind of a turned-on radio pulsar and matter outflowing
from the companion star (see also Stella et al.~2000). Alternatively,
the quiescent X-rays could be due to direct dipole radiation from the
radio pulsar (Di Salvo \& Burderi 2003; Burderi et al. 2003). It is
still unclear why SAX J1808.4--3658 in its quiescent state is
different from the other quiescent neutron-star transients. Its
pulsating nature during outburst distinguishes this source from most
other neutron-star transients and could be related to its unusual
quiescent properties.  More neutron-star transients have to be
observed in quiescence before we can begin to understand SAX
J1808.4--3658.

The four other accreting millisecond pulsars (see Wijnands 2003 for a
review) are prime candidates and three of them will indeed be observed
by either {\it XMM-Newton} or {\it Chandra} in 2004 (see Wijnands et
al.~2004 for the first results). This will clarify whether the unusual
quiescent properties of SAX J1808.4--3658 are related to its pulsating
nature during outburst. In addition, observing and studying in detail
as many non-pulsating neutron-star transients as possible during their
quiescent state will lead to a consensus on the quiescent luminosity
and spectral-shape distribution of those systems. Here we report on a
{\itshape Chandra} observation of the globular cluster Terzan 5 which
is known to harbor the neutron-star transient EXO 1745--248. Our
observation was taken during times when this transient was in its
quiescent state. We found that during its quiescent state the source
exhibited a hard X-ray spectrum similar to SAX J1808.4--3658 although
its 0.5--10 keV luminosity was a factor of $\sim$40 higher than that
of SAX J1808.4--3658. This is the first time that a neutron-star
transient which does not exhibit pulsations during outburst is
observed to exhibit the same spectral behavior as the millisecond
X-ray pulsar SAX J1808.4--3658 in quiescence.

\section{Observations and analysis}

\begin{figure}
\begin{center}
\begin{tabular}{c}
\psfig{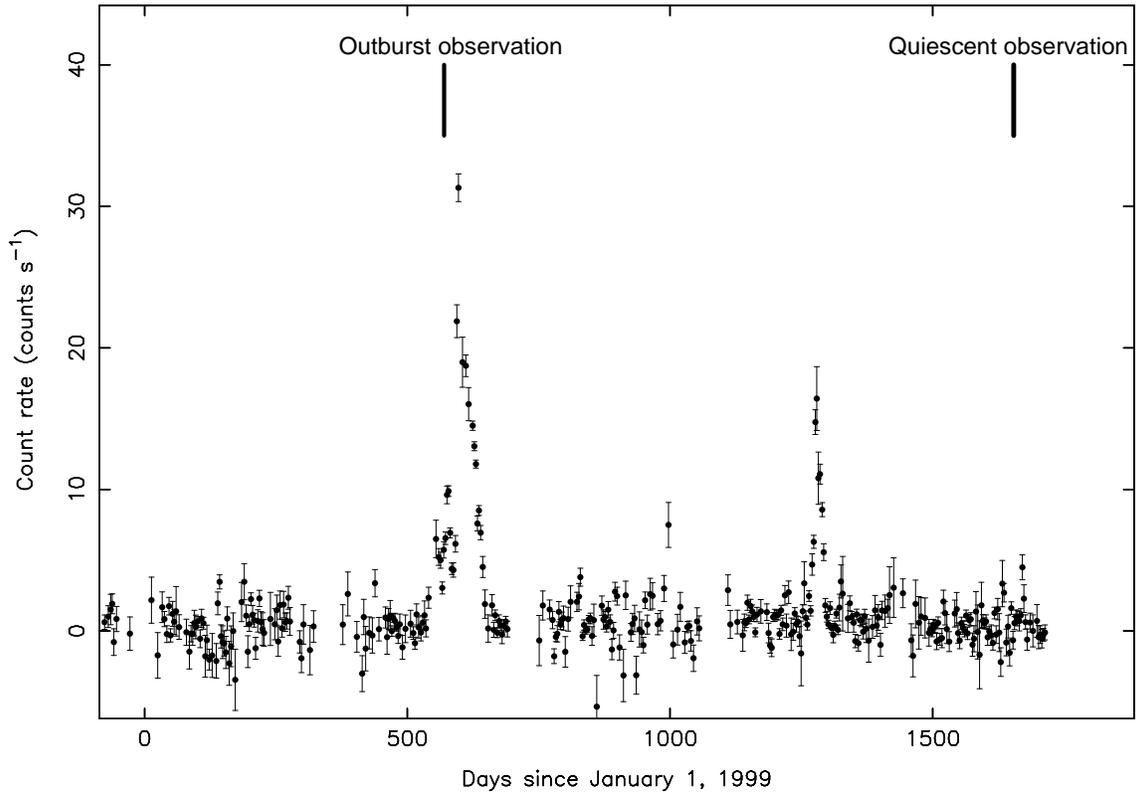}
\end{tabular}
\figcaption{
The {\it RXTE}/ASM light curve of EXO 1745--248. Each point is
averaged over three days. The dates of the {\it Chandra} observations
are marked.
\label{fig:asm} }
\end{center}
\end{figure}

A previous {\itshape Chandra} observation of the globular cluster
Terzan 5 was performed to study the low-luminosity X-ray sources in
this cluster, including the identification and study of the quiescent
counterpart of EXO 1745--248. However, it was found that the
neutron-star transient EXO 1745--248 was in a rare bright state during
this observation (Heinke et al. 2003a; see also
Fig.\ref{fig:asm}). Although this directly resulted in a sub-arcsecond
position for the transient (the errors on the position were
$\sim$0.5$''$; Wijnands, Homan \& Remillard 2002; Heinke et
al. 2003a), its very bright wings (because of the
point-spread-function of the instrument) did not allow Heinke et
al.~(2003a) to study the low-luminosity X-ray sources in detail (only
nine low-luminosity sources could be detected). To obtain the original
goals, we re-observed the cluster using {\it Chandra} on 13--14 July
2003 with a total exposure time of $\sim$39.3 ksec (using the ACIS-S3
chip).  In Figure~\ref{fig:asm} we show the {\it Rossi X-ray Timing
Explorer} ({\itshape RXTE}) All Sky Monitor (ASM) light curve of the
source since January 1999 (part of this light curve was already shown
by Heinke et al. 2003a). In this figure we indicate when our {\it
Chandra} observation was made, showing that at this time EXO 1745--248
was in its quiescent state.  The {\it Chandra} data were reduced and
analyzed using CIAO 3.1 and following the thread listed on the CIAO
web pages\footnote{Available at http://cxc.harvard.edu/ciao/}. We
found that during the final stages of the observation several bright
background flares occurred. These flares were excluded from all
subsequent analyzes resulting in a total usable exposure time of
$\sim$31.2 ksec.

\subsection{Image analysis}

\begin{figure}
\begin{center}
\begin{tabular}{c}
\psfig{figure=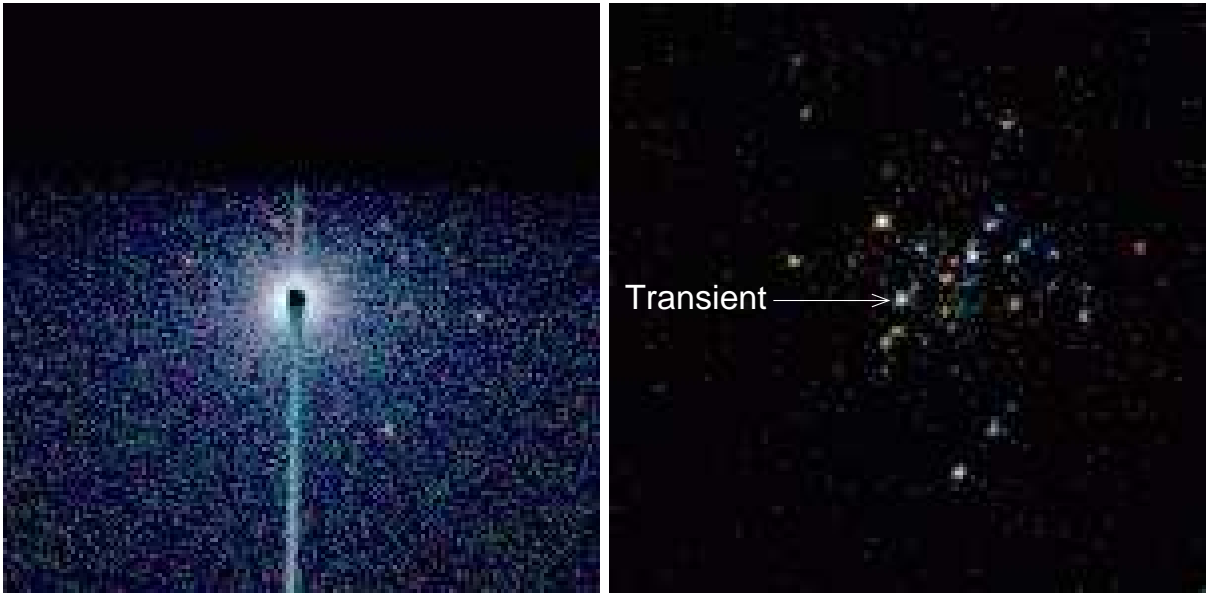,width=16cm}
\end{tabular}
\figcaption{
The images of the globular cluster Terzan 5. The left panel shows the
data obtained in 24 July 2000 during which the transient EXO 1745--248
was very bright (the streak is due to out-of-time events recorded
during the frame transfer); the right panel shows the data obtained on
13--14 July 2003 when the transient was in its quiescent state. Both
images are plotted on the same scale allowing for direct comparison
and are $73.8'' \times 73.8 ''$. East is toward the left and north is
upward.  The red color is for the 0.3--1.5 keV energy range, green for
1.5--2.5 keV, and blue for 2.5--8.0 keV. In the right panel the
quiescent transient is indicated by the arrow. The closest source to
the quiescent transient is approximately 2.4$''$ located toward the
north-west of the source.
\label{fig:color_image} }
\end{center}
\end{figure}

We produced an image of the cluster for the data obtained on 13--14
July 2003 (see Fig.~\ref{fig:color_image} right panel) and for
comparison we made a similar image for the data obtained during the
observation when EXO 1745--248 was in outburst
(Fig.~\ref{fig:color_image} left panel; see also Heinke et
al. 2003a). The images were created after we had reprocessed the data
to remove the pixel randomization added in the standard
processing. This reprocessing slightly enhances the spatial resolution
of the images.  From Figure \ref{fig:color_image} (right) it can be
seen that the transient was indeed observed during its quiescent
state.  We used the tool 'wavdetect' on the quiescent data to search
for low-luminosity point sources and to obtain sub-arcsecond
positions\footnote{In general, the errors on the coordinates of the
detected sources are dominated by the satellite pointing error which
is approximately 0.6$''$ (90\% confidence levels; see
http://cxc.harvard.edu/cal/ASPECT/celmon/).} for those sources. Due to
the crowding in the image, wavdetect did not detect all sources in the
center of the cluster and several sources had to be added by
hand. Including those sources, we detected a total of 61 sources
within two half-mass radii of the cluster. This number of detected
sources is a significant improvement over the ten sources (including
the transient) detected by Heinke et al. (2003a). All the nine
low-luminosity sources detected by Heinke et al.~(2003a) during the
outburst observation were again detected during our quiescent
observation. To identify the individual sources, we use the source
numbering used by Heinke et al.~(2003a) in the remainder of this
paper.  Several of the 61 detected sources are sufficiently bright to
expect that they should have been detected during the outburst
observation if they had been equally bright during both observations,
suggesting that they are highly variable. Moreover, several of the
nine sources detected by Heinke et al. (2003a) are also variable
between the two observations. The analysis and discussion of those
low-luminosity globular cluster sources will be reported in a
forthcoming paper (Heinke et al.~2004 in preparation).

By comparing the outburst and quiescent images, the most likely
counterpart of the transient in its quiescent state can be identified
(see Fig.~\ref{fig:color_image} right; the arrow indicates the likely
quiescent counterpart). Using differential astrometry between the
outburst and quiescent observations, we can determine whether this
likely counterpart candidate is indeed EXO 1745--248 in its quiescent
state. To do that, we calculated the coordinate offset of each
source\footnote{We did not use source eight because it had
significantly larger errors on its coordinates than the other sources
and this source dominated the standard deviation of the averaged
offset. We note that the quiescent data were not tied to any
optical/IR image as was done for the outburst image (Heinke et
al.~2003a). However, for our purposes we are only interested in
differential astrometry between the quiescent and outburst image. The
absolute astrometry of the quiescent image will be discussed by Heinke
et al.~2004, in preparation.} detected by Heinke et al.~(2003a)
between the outburst and quiescent observations. We found an averaged
offset of 0.48$''$ and --0.11$''$ for the right ascension and
declination, respectively. The standard deviations on these offsets
are 0.13$''$ for the right ascension and 0.20$''$ for the
declination. The coordinates given for EXO 1745--248 by Heinke et
al.~(2003a) were offset by 0.48$''$ in right ascension and --0.19$''$
in declination from the (uncorrected) coordinates of the candidate
quiescent counterpart (indicated by the arrow in
Fig.~\ref{fig:color_image}, right panel). These offsets are well
within the range of offsets measured for the other sources.

From the above, we see that the two X-ray images can be tied to each
other to within $\sim$0.2$''$ (i.e., the standard deviations on the
offsets). We can now estimate the likelihood for a random source to
fall within 0.2$''$ of the source position of EXO 1745--248. We
calculated this probability using the source density in one core
radius of the globular cluster (which is 7.9$''$; Cohn et
al.~2002). We detected 13 sources in this core radius which gives a
probability of 0.8\% that one of those 13 sources fall within 0.2$''$
of the position of EXO 1745--248. We note that this probability is
conservative since the source density increases toward the cluster
center and EXO 1745--248 is located close to the edge of the core
where the source density is less. Moreover, the source detected on the
position of EXO 1745--248 has unusual properties for a low-luminosity
globular cluster X-ray source: it has a relatively high X-ray
luminosity and hard X-ray spectrum (see
\S~\ref{section:spectral}). Those sources are very unusual in globular
clusters and the chances for such a peculiar source (of which there
are two within the core radius of Terzan 5; see
Fig.~\ref{fig:color_image} and Heinke et al.~2004 in preparation) to
fall within 0.2$''$ of the position of EXO 1745--248 is
0.12\%. Therefore, we conclude that this low-luminosity source is
indeed the quiescent counterpart of the transient EXO 1745--248 and
that it has peculiar characteristics in its quiescent state (as
discussed below).

\subsection{Spectral analysis \label{section:spectral}}

From Figure~\ref{fig:color_image} (right) it can be seen that the
quiescent counterpart was not the brightest X-ray source in the
cluster and that it had somewhat blueish X-ray colors indicating a
reasonably hard spectrum.  To quantify this, we extracted the
quiescent spectrum of the source. We used the randomized data to
extract the spectrum since no calibration products are available for
data with the randomization removed. We used a circle with a radius of
1.5$''$ centered on the source position as the extraction region.
Owing to the crowding near the position of the transient, the standard
practice of using an annulus around the source position as extraction
region for the background could not be applied. Therefore, as
background region we used a circle with a radius of 40$''$ which was
1.3$'$ to the west of EXO 1745--248 but which did not contain any
point sources. We used the CIAO tool 'psextract' to extract the source
and background spectra and to create the response matrix and the
ancillary response files (the latter was automatically corrected for
the time-variable low-energy quantum efficiency
degradation\footnote{See
http://asc.harvard.edu/cal/Acis/Cal\_prods/qeDeg/}). We grouped the
spectrum in bins of 15 counts in order to validate the use of the
$\chi^2$ fitting method and fitted the spectrum using Xspec (Arnaud
1996; V11.3.1).

\begin{figure}
\begin{center}
\begin{tabular}{c}
\psfig{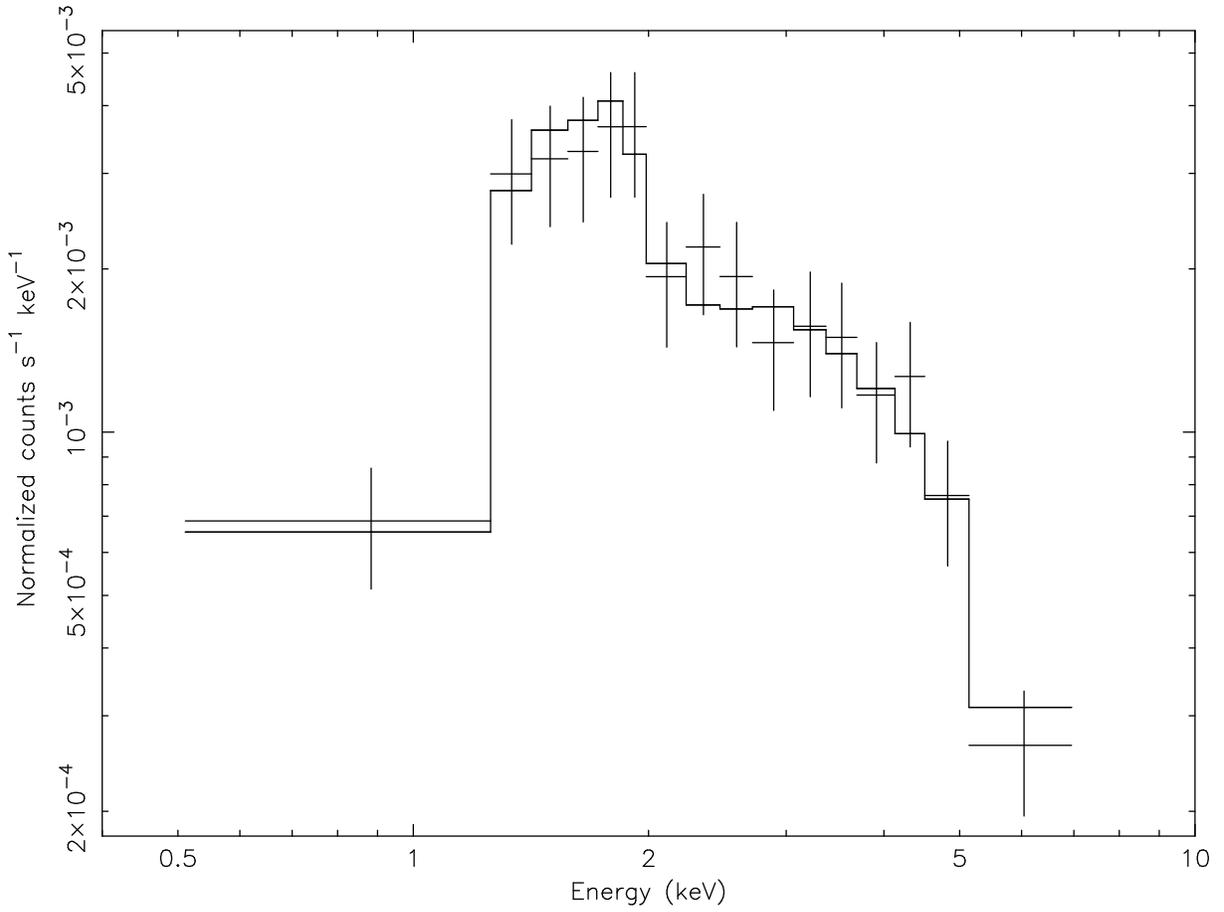}
\end{tabular}
\figcaption{
The quiescent X-ray spectrum obtained for EXO 1745--248 during the
13--14 July 2003 {\itshape Chandra} observation. The solid line
indicates the best fit power-law model.
\label{fig:spectrum} }
\end{center}
\end{figure}

The source spectrum is shown in Figure \ref{fig:spectrum} (a total of
$\sim$250 counts were detected from EXO 1745--248; see
\S~\ref{section:timing}). Clearly, the source is significantly
detected up to at least 6 keV confirming the hardness of the
spectrum. We fitted the spectrum using a variety of models. We found
that a simple power-law model modified by the Galactic absorption
column density $N_{\rm H}$ (using 'phabs' in Xspec; the column density
was left as a free parameter in the fits) toward Terzan 5 could
adequately fit the data (reduced $\chi^2$ of 0.29 with 13 degrees of
freedom [d.o.f.]\footnote{The reduced $\chi^2$ is rather low and might
suggest that the $\chi^2$ method cannot be used for our data set. We
have checked our results by using different number of counts per
spectral bin, by using the Cash statistics (Cash 1979), and by using
both. The results obtained with those alternative methods were always
consistent with those obtained with the $\chi^2$ method using 15
counts per bin (i.e., the fractional uncertainties obtained when
calculating errors using Cash statistics are very similar to those
measured when using $\chi^2$ statistics). Therefore, we only present
the results obtained via the latter method.}). The obtained $N_{\rm
H}$ was 1.4$^{+0.5}_{-0.4}\times 10^{22}$ cm$^{-2}$ which is
consistent with the infrared-derived estimate for the column density
of Cohn et al. (2002; who obtained $1.2\times 10^{22}$
cm$^{-2}$)\footnote{During outburst, Kuulkers et al. 2003 reported
column densities which could be up to three times larger than that
measured for the quiescent counterpart of EXO 1745--248. However,
during outburst EXO 1745--248 can exhibit strong dipping behavior
(e.g., Markwardt \& Swank 2000) suggesting that those measured column
densities are due to internal absorption in the system. Furthermore,
Heinke et al.~2003a reported on a {\it Chandra} outburst observation
of this source and they measured a column density which is more
consistent with the infrared-derived estimate (albeit still slightly
larger) also suggesting variable internal
absorption. \label{footnote_nh}}. The power-law index obtained was
1.8$^{+0.5}_{-0.4}$ and the unabsorbed 0.5--10 keV flux was
$2.2^{+0.7}_{-0.3} \times 10^{-13}$ \funit. Normally quiescent
neutron-star systems are fit with a thermal model, i.e., a
neutron-star atmosphere (NSA) model (Zavlin, Pavlov, \& Shibanov 1996,
G\"ansick, Braje, \& Romani 2002). In those NSA models, the
normalization is given as $1/d^2$ with $d$ the distance in pc. The
distance toward Terzan 5 is given as 8.7\pp3 kpc (Cohn et al. 2002;
Kuulkers et al. 2003), resulting in a $1/d^2$ of $1.32\times 10^{-8}$,
which was used as normalization in our initial fits using the two NSA
models. We also assumed a 'canonical' neutron star with a mass of 1.4
\mzon~and a radius of 10 km. Therefore, in the NSA models
only the temperature was allowed to float as a free
parameter. However, when fitting the quiescent spectrum, we found that
such models did not provide a good fit with a reduced $\chi^2>5.3$
(using 14 d.o.f.). Leaving the neutron-star mass and/or radius free in
the fits also did not result in an acceptable description of our
spectrum (reduced $\chi^2>3.3$; 12 d.o.f.) Furthermore, when leaving
those parameters free in the fits, they could not be constrained.

Leaving the normalization free and the column density as a free
parameter (but again assuming a canonical neutron star) resulted in an
acceptable fit for the NSA model from Zavlin et al.~(1996) but with an
effective temperature ($T_{\rm eff}$) of $\sim$0.6 keV (for an
observer at infinity) and a normalization of $\sim4.2\times
10^{-12}$. This temperature is a factor of $\sim$3 larger than usually
seen for quiescent neutron-star systems and the normalization implies
an extremely large distance toward the source of $\sim$490
kpc. Therefore, we conclude that this NSA model does not provide a
good description of the data. The NSA model of G\"ansick et al. (2002)
did not allow for a good fit because the temperature soon reached the
maximum allowed temperature in that model ($\log T_{\rm eff} = 6.5$)
suggesting that also this NSA model does not describe our data. For
comparison with results in the literature for other quiescent
neutron-star transients, we also fit the data with a black-body model,
but the resulting temperature was 1.0$\pm$0.2 keV, again significantly
larger than seen in any other quiescent neutron-star system (the
resulting column density was 0.6$\pm$0.3 cm$^{-2}$ and the reduced
$\chi^2$ was 0.36 with 12 d.o.f.). Furthermore, the obtained radius of
the emitting region was very small (only 0.10$\pm$0.04 km; assuming a
distance of 8.7 kpc) also suggesting that a black-body model is not a
good description of our data.

Several quiescent neutron-star transients have a quiescent spectrum
which consists of a soft, thermal component and an additional hard
component which dominates above a few keV.  One can assume that EXO
1745--248 has a similar composite quiescent spectrum and that the
relatively high column density toward Terzan 5 obscures most (if not
all) of the thermal component. To determine the flux upper limit on
such a soft component we fitted the spectrum using the NSA models plus
a power-law component.  We again assumed a canonical neutron star and
fixed the normalization in the NSA models to $1/d^2 =
1.32\times10^{-8}$. The only free parameter of the NSA models was
again the temperature. Such models resulted in a best fit value for
$T_{\rm eff}$ (for an observer at infinity) of $<0.08$ keV with an
unabsorbed 0.5--10 keV flux of $0.1-0.2\times 10^{-13}$ \funit~for
both the Zavlin et al. (1996) and G\"ansick et al. (2002) NSA models
(the bolometric flux was approximately twice these values; the
power-law index and column density were consistent with that obtained
when fitting the data with only a power-law component). These upper
limits on the effective temperatures are close to the temperatures
observed for other quiescent neutron-star transients, however, the
contribution of the thermal component to the unabsorbed 0.5--10 keV
flux is only 5\%--7\%. Changing the normalization (hence the distance;
we used a range between 6 and 12 kpc) only resulted in a maximum
contribution of the thermal component to the unabsorbed 0.5--10 keV
flux of 8\%--10\%, strongly suggesting that the quiescent spectrum of
EXO 1745--248 is more than 90\% dominated by the hard spectral
component.

A different way of demonstrating that the quiescent spectrum of EXO
1745--248 is dominated by the hard component is assuming that both the
thermal and hard component contribute $\sim$50\% to the unabsorbed
flux in the 0.5--10 keV energy range. We fixed the normalization of
the power-law component in such a way that this component always
contributed 50\% to the 0.5--10 keV flux. This time we allowed both
the temperature of the NSA models and the normalization to be free
parameters during the fits (again we assumed a canonical neutron
star). This resulted in relatively high effective temperatures
($T_{\rm eff}$ for an observer at infinity of $>0.3$ keV) but with
unrealistically large distances of up to several Mpc.  Therefore, such
models do not provide a realistic description of the spectrum,
suggesting again that the spectrum is dominated by the hard power-law
component with only a small contribution from the thermal component.

Finally, we fitted the spectrum with a thermal bremsstrahlung spectrum
which could fit the data satisfactorily (reduced $\chi^2$ of 0.24 with
12 d.o.f.). The column density obtained was $1.2\pm0.4\times10^{22}$
cm$^{-2}$, again very similar to the inferred column density toward
Terzan 5, and the plasma temperature was 8$^{+17}_{-4}$ keV. The
unabsorbed 0.5--10 keV flux was $2.0\times 10^{-13}$ \funit.

\subsection{Timing analysis \label{section:timing}}

\begin{figure}
\begin{center}
\begin{tabular}{c}
\psfig{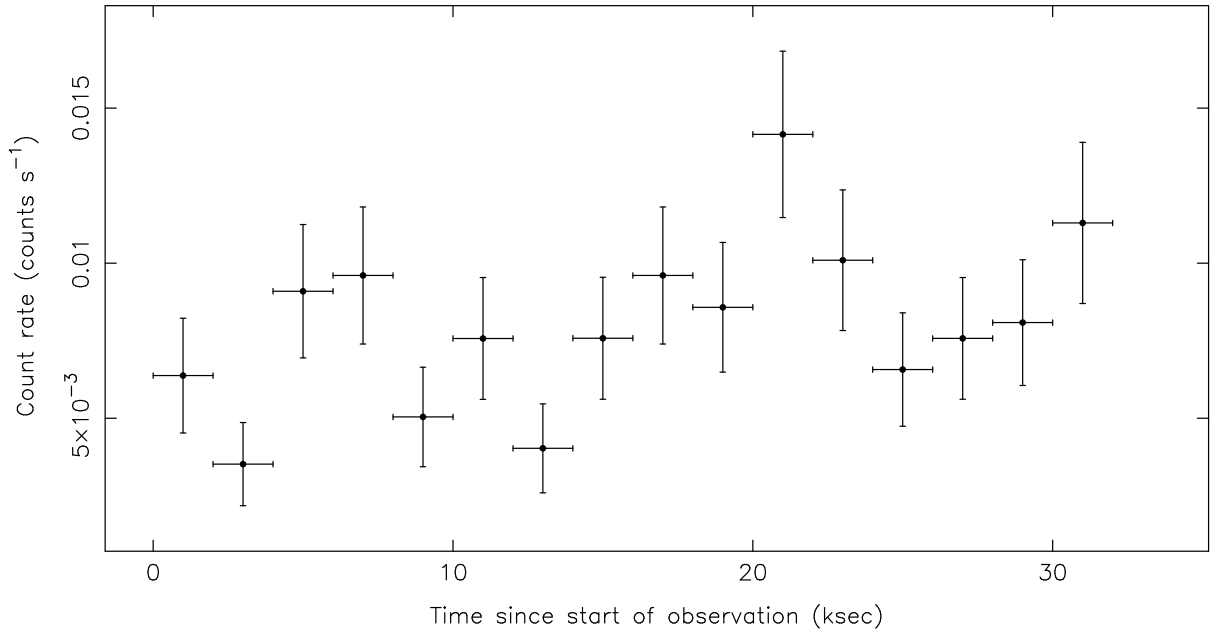}
\end{tabular}
\figcaption{
The X-ray count rate curve (0.3--10 keV) for EXO 1745--248 during the
13--14 July 2003 {\itshape Chandra} observation. The time resolution
is 2000 seconds.
\label{fig:lc} }
\end{center}
\end{figure}

The number of counts detected for EXO 1745--248 (using the same source
and background regions as in the spectral analysis above) was 250\pp16
counts (0.3--10 keV), resulting in a background-corrected count rate
of 0.0080\pp0.0005 counts s$^{-1}$ (averaged over the entire
observation). To determine the amount of variability in the count rate
throughout the observation, we created X-ray count rate curves of the
source. In Figure \ref{fig:lc} the count rate curve is shown using a
time resolution of 2000 seconds. This figure shows that the source was
variable during our observation. We applied Kolmogorov-Smirnov and
Cramer-Von Mises tests (using the method outlined in Heinke et
al. 2003a) on the 0.5--8 keV event list of the source to attempt to
disprove the hypothesis that the source flux is constant. Both tests
show that EXO 1745--248 is variable at the 95\% confidence level.

\section{Discussion}

We have presented a {\itshape Chandra} observation of the globular
cluster Terzan 5 during times when the neutron-star X-ray transient
EXO 1745--248, known to be a member of this cluster, was in its
quiescent state. The transient was detected at a 0.5--10 keV
luminosity of $\sim2 \times 10^{33}$ \Lunit~for a distance of 8.7 kpc
(Cohn et al. 2002; Kuulkers et al. 2003). This quiescent luminosity of
EXO 1745--248 is similar to that observed for several quiescent
neutron-star systems (e.g., 4U 1608--52; Aql X-1; 4U 2129+47; Asai et
al. 1998; Rutledge et al. 2001b; Nowak et al. 2002). However, in
contrast to those systems, the quiescent spectrum of EXO 1745--248 is
not dominated by the thermal component but by the hard spectral
component. We estimate that the soft component contributes at most
5\%--10\% (depending on the exact spectral model used and the
uncertainties in the distance) of the flux in the 0.5--10 keV energy
range (thus a luminosity upper limit of $1-2\times 10^{32}$ \Lunit)
and that the temperature of the thermal component is very low ($<0.1$
keV). The source is also variable at the 95\% confidence level.

\subsection{Thermal emission from the neutron star in EXO 1745--248}

The standard cooling model (which assumes standard core cooling
processes; see \S~\ref{section:introduction}) is commonly used to
explain the thermal emission of quiescent neutron-star systems. In
this model, the expected quiescent flux ($F_{\rm q}$) of a particular
system depends on its long-term time-averaged (averaged over $>10,000$
years) accretion history and is given by $F_{\rm q} \approx \langle F
\rangle/135 $ (Wijnands et al. 2001; Rutledge et al. 2002) with
$\langle F \rangle$ the time averaged flux due to accretion. The
latter can be rewritten as $\langle F \rangle = t_{\rm o} \langle
F_{\rm o} \rangle / (t_{\rm o} + t_{\rm q})$ resulting in $F_{\rm q}
\approx {t_{\rm o} \over t_{\rm o} + t_{\rm q}} \times {\langle F_{\rm
o} \rangle \over 135}$, with $\langle F_{\rm o} \rangle$ the average
flux during outburst, $t_{\rm o}$ the average time the source is in
outburst, and $t_{\rm q}$ the average time the source is in
quiescence.  EXO 1745--248 has been detected in outburst on several
occasions\footnote{Due to the limited angular resolution of past
instruments, it cannot be excluded that another X-ray transient in
Terzan 5 was detected during those observations instead of EXO
1745--248. However, only one globular cluster (M15) has so far been
found to harbor two bright X-ray binaries (White \& Angelini 2001),
making it more likely that the bright transient in Terzan 5 was always
EXO 1745--248.}: August 1980 (Makishima et al. 1981; type-I X-ray
bursts were seen indicating active accretion onto the neutron star but
no persistent emission could be found), June 1984 (Warwick et
al. 1988), August 1990 (Verbunt et al. 1995), March 1991 (Johnston,
Verbunt, \& Hasinger 1995), July-August 2000 (Markwardt \& Swank 2000;
Kuulkers et al. 2003), and June 2002 (Wijnands et al. 2002). The
observations of the source obtained during all outbursts before the
2000 outburst constitute only a snap-shot of the brightness of the
source. It is unknown how long those outbursts lasted and how bright
the source became during each of them.  The {\it RXTE}/ASM light curve
(see Fig.~\ref{fig:asm}) is available only for the 2000 and 2002
outbursts and can be used to provide an estimate of the time-averaged
accretion flux over the last decade. We calculated the expected
quiescent flux of EXO 1745--248 mostly based on what is observed
during the 2000 and 2002 outbursts.

During the 2000 and 2002 outbursts, the source was detected for
$\sim$109 and $\sim$30 days, respectively, with a corresponding
time-averaged ASM count rate of $\sim$9.7 and $\sim$7.9
counts~s$^{-1}$. No other outbursts were observed during the active
lifetime of {\it RXTE}/ASM (which has been active for 3076 days; as of
July 23, 2004). Therefore, averaged over the two observed outbursts,
the source was active for 139 days with a time-averaged ASM count rate
during outburst of 9.3 count s$^{-1}$. To calculate the corresponding
time-averaged flux of the source, we used PIMMS\footnote{Available at
http://heasarc.gsfc.nasa.gov/Tools/w3pimms.html} to convert this count
rate into a flux estimate assuming a power-law shaped spectrum with a
power-law index of 2. During our quiescent observation, we have found
an $N_{\rm H}$ of $\sim1.3 \times 10^{22}$ cm$^{-2}$ toward EXO
1745--248, however when the source was in outburst, the observed
$N_{\rm H}$ could be up to three times larger (Kuulkers et al. 2003;
likely due to internal absorption in the system; see
footnote~\ref{footnote_nh}). Therefore, we assumed a range of column
density between 1 and $3 \times 10^{22}$ cm$^{-2}$ when using PIMMS.
This resulted in a time-averaged flux (0.1--100 keV) during the
outbursts of $1.2 - 1.5 \times 10^{-8}$ \funit. During the active
life-time of the {\it RXTE}/ASM the source was active for 139 days and
quiescent for 2937 days (as of July 23, 2004) resulting in a
time-averaged flux for the life-time of the ASM of $5.4 - 6.8 \times
10^{-10}$ \funit. Since the source has likely spent more time in
quiescence before the launch of {\it RXTE}, these fluxes should be
considered upper-limits on the true time-averaged accretion flux. The
outburst history before {\it RXTE} is not well known but to first
approximation we can assume that no outbursts occurred between the
detection of the source in March 1991 (Johnston et al. 1995) and the
2000 outbursts. If true, then the time-averaged source flux is $3.4 -
4.3 \times 10^{-10}$ \funit~for the time period between the 1991
detection of the source and the present-day. Because one or more
outbursts might have been missed, these flux estimates should be
regarded as lower-limits. Combining the two limits, the time-averaged
flux of the source due to accretion is likely in the range 3.4 to $6.8
\times 10^{-10}$ \funit, resulting in a predicted quiescent thermal
flux of $2.5 - 5.0 \times 10^{-12}$ \funit~(if only standard core
cooling is taken into account). However, these predicted values are a
factor of $\sim$100 larger than the flux upper limit we determined for
the thermal component in the quiescent data of EXO 1745--248,
suggesting that the neutron star is colder than expected, likely owing
to the presence of enhanced core cooling processes.

We note that this calculation of the predicted thermal quiescent flux
is based on very limited information about the accretion history of
the source.  Furthermore, we have assumed a simple power-law shaped
spectrum of the source during outburst but it has already been shown
that this is an oversimplification of the true situation (see, e.g.,
Kuulkers et al. 2003; Heinke et al. 2003a), resulting in additional
uncertainties in the predicted quiescent flux.  However, in order for
the standard cooling model to agree with our observations, the
predicted quiescent flux must drop by a factor of 100, meaning that
the quiescent episodes between very bright outbursts (like the 2000
and 2002 ones) should be over a thousand years. This seems unlikely
since the source has been detected on several occasions before the two
latest outbursts. We prefer the point of view in which the low thermal
emission from the neutron star is explained by assuming that enhanced
cooling processes occur in the core, especially since strong evidence
has been obtained for such processes in the neutron stars of several
other quiescent systems (e.g., Brown et al. 1998; Wijnands et
al. 2001, 2003; Nowak et al. 2002; Campana et al. 2002).  Colpi et
al. (2001) proposed that the presence of enhanced core cooling
processes indicates that the neutron star should be relatively massive
($>1.7$ \mzon).  We note that this conclusion remains valid even if
the low-luminosity source we have detected at the position of EXO
1745--248 is not the quiescent counterpart of this transient (although
as explained above this is unlikely) but in fact an unrelated source
(also likely a member of the globular cluster).

\subsection{The hard spectral component of EXO 1745--248}

The fact that EXO 1745--248 might harbor a cold neutron star may
explain why the quiescent spectrum of EXO 1745--248 is dominated by
the hard spectral component. Several other systems have exhibited hard
components in quiescence which contributed up to 40\% to the 0.5--10
keV quiescent flux (\S~\ref{section:introduction}).  Wijnands et
al. (2003) suggested that if a system was found which has a similar
luminous hard component as seen, e.g., for Aql X-1 in its quiescent
state (e.g., Asai et al. 1998), but with a neutron star which only
emitted very faint thermal emission (e.g., because of its coldness),
then the quiescent spectrum of this system should be dominated by the
hard spectral component. It seems that EXO 1745--248 might indeed be
such a system. We note that this argument assumes that the thermal and
hard components are not related to each other and can therefore vary
in strength independently. It also assumes that the hard component in
EXO 1745--248 is due to the same process which causes the hard
components observed in other systems. It is unclear how valid those
assumptions are, mainly because of the lack of understanding of the
hard X-ray emission seen in quiescent neutron-star X-ray transients.

It has been suggested that this hard component might be due to
residual accretion onto the surface or magnetic field of the neutron
star or to the pulsar mechanism being active (see the discussion in
Campana \& Stella 2000). Only limited investigations have been
performed to determine exactly how those mechanisms could produce the
hard emission, what exactly the spectral shape of the emission would
be, and which source properties would make sources differ from each
other with respect to this hard spectral component.  Currently only
two neutron-star systems are known whose quiescent spectra are
dominated by the hard spectral component: EXO 1745--248 and SAX
J1808.4--3658. Campana et al. (2002) found that SAX J1808.4--3658 is
dominated by the hard component in quiescence, although the luminosity
of SAX J1808.4--3658 was a factor of $\sim$40 lower than that of EXO
1745--246. Again it is unclear whether or not the hard components in
these sources are due to the same phenomenon. However, despite this
uncertainty, it is clear that hard quiescent spectra can occur at
relatively high and very low luminosities and presumably also for
luminosities in-between those observed for EXO 1745--248 and SAX
J1808.4--3658. Our results also demonstrate that it is not necessary
for a source to be an accretion-driven millisecond X-ray pulsar during
outburst for its quiescent spectrum to be dominated by the hard
component. For EXO 1745--248 no pulsations were seen during the
outbursts indicating a different magnetic field structure (e.g.,
strength, configuration, multi-pole moments) for the neutron star in
this system than for the neutron star in SAX J1808.4--3658.  However,
several models for the hard spectral component involve a combination
of the magnetic field strength and configuration, and residual
accretion from the companion star, all unknown properties. This gives
a large range of source properties one can vary to obtain similar
quiescent spectral shapes but different outburst behavior. Insight
into the physical process(es) behind the hard spectral component might
come from the variability in this component observed in EXO 1745--248
(see Fig.~\ref{fig:lc}).

The hard quiescent spectra of EXO 1745--248 and SAX J1808.4--3658
(Campana et al. 2002) are very similar to the hard quiescent spectra
observed for the quiescent black-hole X-ray transients (see Kong et
al. 2002 and references therein). It is not clear if the hard spectral
component is caused by the same physical process(es) in the quiescent
neutron-star transients as in the quiescent black-hole systems, but if
they are then any models involving a neutron-star surface or magnetic
field or a black-hole event horizon are not valid.  Irrespective of
whether or not the same physical mechanism(s) is behind the hard
quiescent spectra of EXO 1745--248 and SAX J1808.4--3658 and the
black-hole systems, our results and those of Campana et al. (2002)
demonstrate that not all the neutron-star systems are different in
quiescence from the black-hole systems. Rutledge et al. (2000)
suggested that these differences could be used to determine the nature
of the compact object in an X-ray binary for which this was not yet
known. However, our results and those of Campana et al. (2002) now
show that a hard spectral shape in quiescence does not mean that the
compact object in a particular X-ray transient is a black hole (note
that on average, the neutron-star systems could still be different
from the black-hole systems, but one particular neutron-star binary
might not be). On the other hand, the faintness of several of the
black-hole systems might still be a property potentially exclusive to
those systems and might be used to determine the black-hole nature of
certain transients. Unfortunately, several quiescent black-hole
systems have quiescent luminosities in the range seen for SAX
J1808.4--3658 and EXO 1745--248, making this possible method of
distinguishing between the different types of transients only useful
for those black-hole systems which can become very faint in
quiescence. If the quiescent spectrum of a particular transient is
dominated by a soft thermal component, it is still likely that this
system harbors a neutron-star primary since no black-hole system has
shown a thermal component in its quiescent X-ray spectrum.

Based on the outburst spectrum of EXO 1745--248, Heinke et al. (2003a)
suggested that EXO 1745--248 might be in an ultra-compact binary with
a binary period $<$ 1 hour. The 2-hr binary period of SAX
J1808.4--3658 (Chakrabarty \& Morgan 1998) shows that this system is
in a compact binary system. If the compact binary nature of EXO
1745--248 can be confirmed, this would tentatively indicate that the
compact binary nature of this source and SAX J1808.4--3658 might be
related to their hard spectra in their quiescent states. However,
currently not enough information is available to arrive at any
definite conclusions. More information will be available in the next
few years after more (ultra-)compact neutron-star binaries have been
observed and studied in quiescence (such as XTE J1751--305, XTE
J0929-314, and XTE J1807--294 which are currently scheduled to be
observed with {\it Chandra} and/or {\it XMM-Newton}; see Wijnands et
al.~2004 for the first results).

\subsection{Implication for the classification of low-luminosity 
globular cluster sources}

\begin{figure}
\begin{center}
\begin{tabular}{c}
\psfig{figure=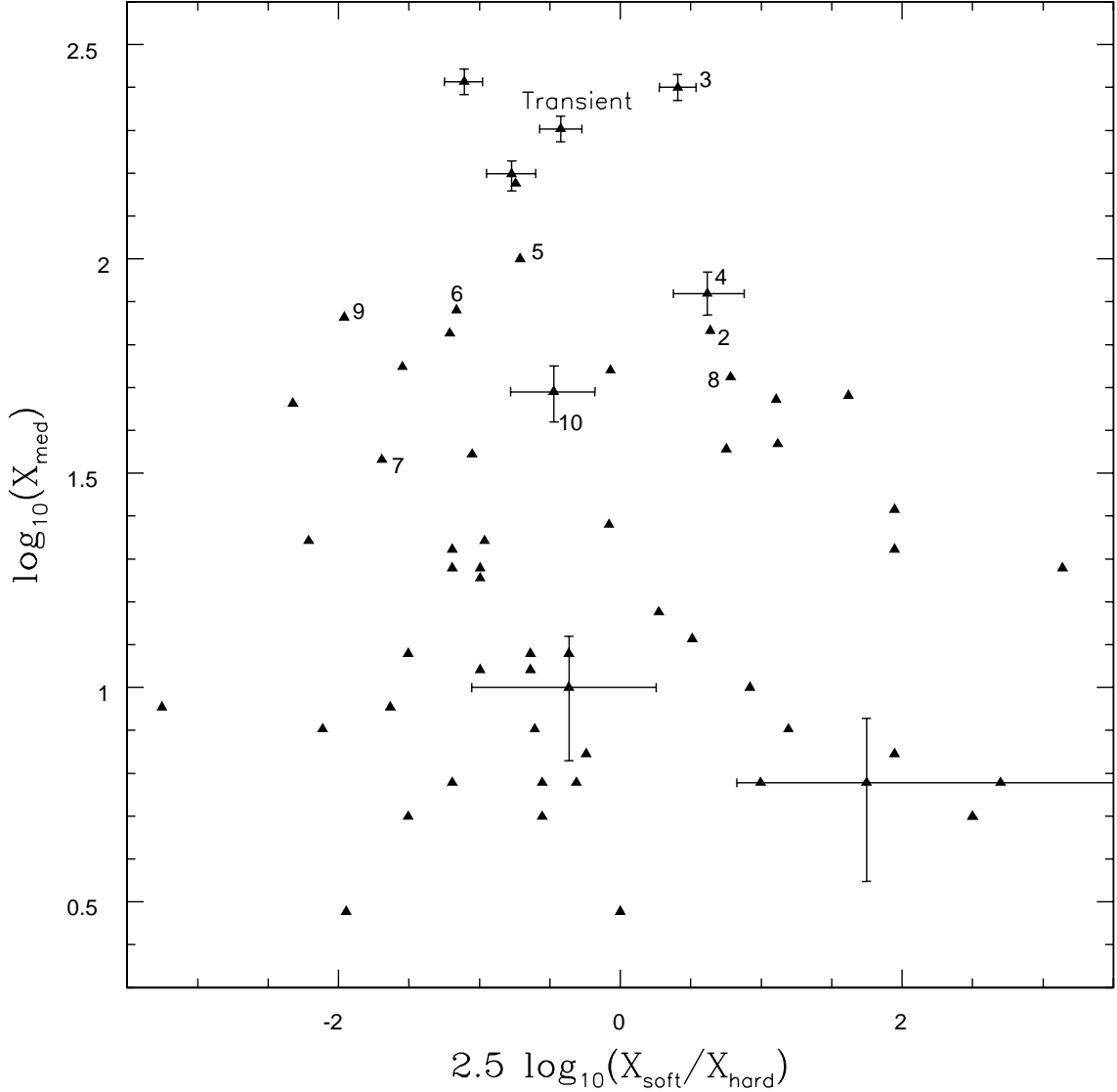,width=16cm}
\end{tabular}
\figcaption{
The color-magnitude diagram of Terzan 5. The color is defined as the
2.5 log$_{10}$ of the ratio of the number of soft photons X$_{\rm
soft}$ (for the energy range 0.5--2 keV) to the number of hard photons
X$_{\rm hard}$ (for the energy range 2--6 keV). The magnitude is
defined as the log$_{10}$ of the number of medium hard photons X$_{\rm
med}$ (for the energy range 0.5--4.5 keV). In the figure the quiescent
neutron-star transient EXO 1745--248 is indicated by the word
'Transient'. The numbers near some of the points correspond to the
low-luminosity sources already detected by Heinke et al. (2003) and
follows their numbering. For clarity only a few representative error
bars are plotted.
\label{fig:cmd} }

\end{center}
\end{figure}

The fact that EXO 1745--248 and SAX J1808.4--3658 have hard quiescent
spectra may have consequences for our understanding of the nature of
the low-luminosity globular cluster X-ray sources. When optical and
radio identifications are not available for those sources, they are
often given tentative classifications on the basis of their X-ray
luminosities and the hardness of their spectra (e.g., Grindlay et
al. 2001a, b; Pooley et al. 2002a, b; Gendre, Barret, \& Web 2003;
Heinke et al. 2003a). Those sources which have luminosities above
$10^{32}$ \Lunit~and are soft are generally classified as quiescent
neutron-stars systems; those which are relatively bright (up to around
$10^{33}$ \Lunit) and have hard spectra are generally classified as
cataclysmic variables (CVs). The sources below $\sim10^{31}$
\Lunit~could be a variety of objects, including CVs, millisecond radio
pulsars or active binaries (e.g., BY Dra or RS CVn systems). Usually a
X-ray color-magnitude diagram is created and different branches are
identified corresponding to different source types. To investigate the
impact of the hard spectrum of EXO 1745--248 on such a diagram we
create a similar diagram for Terzan 5 (Fig.~\ref{fig:cmd}). Clearly,
the quiescent transient (labeled 'Transient') is one of the brightest
sources and its color is quite hard.  The sources which are identified
by the numbers 2, 3, 4, and 8 (using the numbering of Heinke et
al. 2003a) are considerably softer and would likely be classified as
quiescent neutron-star systems if no additional information were
available. Indeed Heinke et al. (2003a) classified them as such and
they suggested that the sources to the left of them were good CV
candidates. Without prior knowledge of the nature of EXO 1745--248,
its hard color would have suggested it to be a CV, but this is far
from a clear-cut classification. Moreover, when candidate CVs have
enough counts to extract their X-ray spectra, those spectra are
usually fit with a thermal bremsstrahlung model with a plasma
temperature of $\ga$ 5 keV. As shown in \S~\ref{section:spectral}, the
quiescent spectrum of EXO 1745--248 could also be adequately fitted
with such a spectral model and the resulting plasma temperature is
fully consistent with what has been observed from CVs. The only
unusual aspect of EXO 1745--248 would be its rather high X-ray
luminosity of $\sim 2 \times 10^{33}$ \Lunit, which would not be fully
consistent with a CV nature.  Likely, the source would then have been
classified as an unusually bright CV, but a potential quiescent
neutron-star transient with a very unusual spectrum might also have
been suggested.

This possible misclassification strongly suggests that some of the
hard sources in the globular cluster studies might have been
misclassified as CVs but are in reality quiescent neutron-star
systems. This conclusion is further strengthened by the quiescent
properties of SAX J1808.4--3658 (Campana et al. 2002). This source had
a similar hard quiescent spectrum but a $\sim$40 times lower X-ray
luminosity bringing this source right in the range of luminosity
expected for CVs. It is very likely that if SAX J1808.4--3658 had been
located in a globular cluster and the neutron-star nature of its
primary was not known from additional information (like its outbursts
and pulsations), then this source would have been classified on the
basis of its quiescent X-ray properties as a CV (see also the
discussion in Wijnands et al. 2003). We see no reason why future
discovered quiescent neutron-star systems would not exhibit similarly
hard spectral shapes and have X-ray luminosities in-between what we
observe for EXO 1745--248 and SAX J1808.4--3658. If true, then this
group of sources would have very similar X-ray characteristics to
those of CVs and this must be taken into account during the
classification of low-luminosity X-ray sources in globular clusters
when this is done solely on the basis of their X-ray properties.

We searched the literature and found eleven detections of quiescent
neutron-star systems for which spectra of sufficient quality could be
obtained to investigate their quiescent spectra. We only discuss those
systems which are confirmed neutron-star X-ray transients as they have
been seen to exhibit type-I X-ray bursts and/or X-ray pulsations. Out
of those eleven sources, nine were dominated by the thermal component
and two were dominated by the hard component\footnote{The
soft-component dominated systems are Aql X-1, Cen X-4, 4U 1608--52, 4U
2129+47, KS 1731--260, MXB 1659--29, SAX J1748.9--2021 in the globular
cluster NGC 6440, RX J170930.2--263927, and SAX J1810.8--2609; the
hard-component dominated systems are SAX J1808.4--3658 and EXO
1745--248}. This suggests that out of every eleven quiescent
neutron-star systems two such systems could have hard quiescent
spectra, which is a non-negligible fraction of the total.  Therefore,
a similar fraction of hard quiescent neutron-star transients might be
present in globular clusters which are currently missed because they
are incorrectly classified. We note that this is a very rough estimate
mainly because of the very few systems known 
(especially those with a
hard quiescent spectrum).  This estimate assumes that most accreting
neutron-star systems are similar to those that have been detected in
outburst.  If the spectral shape in quiescence is related to recent
outburst activity, the fraction of hard quiescent neutron-star
transients may be larger or smaller than our calculation suggests.
Heinke et al. (2003b) compiled a list of 21 quiescent neutron-star
systems in globular clusters, identified by their agreement with an
NSA model plus an optional power-law component making up $<40$\% of
the 0.5--10 keV flux.  Their results suggest that the strength of a
power-law component is correlated with recent outburst activity.
However, since their method of identifying quiescent neutron-star
systems would not detect all known systems, that correlation cannot
yet be considered secure.  Reliable optical identifications of
numerous hard X-ray sources with cataclysmic variables in 47 Tuc, NGC
6397, and NGC 6752 (Edmonds et al. 2003; Grindlay et al. 2001b; Pooley
et al. 2002a) indicate that quiescent neutron-star systems with hard
spectra probably do not dominate the quiescent neutron star population
in globular clusters.

\acknowledgments

We used results provided by the {\it RXTE}/ASM teams at MIT and at the
{\it RXTE} SOF and GOF at NASA's GSFC. We especially thank Ron
Remillard for reprocessing the {\it RXTE}/ASM data to obtain the
complete ASM light curve for EXO 1745--248. This research has made use
of NASA's Astrophysics Data System and the tools provided by the High
Energy Astrophysics Science Archive Research Center (HEASARC),
provided by NASA's Goddard Space Flight Center.

\clearpage

\clearpage

\clearpage

\clearpage

\clearpage


\begin{references}

          
\reference{}Arnaud, K. 1996, in G. Jacoby \& J. Barnes (eds.), {\it
Astronomical Data Analysis Software and Systems V.}, Vol. 101, p. 17,
ASP Conf. Series.
                 
\reference{}Asai, K., Dotani, T., Kunieda, H., Kawai, N. 1996, \pasj,
48, L27

\reference{}Asai, K., Dotani, T., Hoshi, R., Tanaka, Y., Robinson,
C. R., Terada, K. 1998, \pasj, 50, 611
                 
\reference{}Brown, E. F., Bildsten, L., \& Rutledge, R. E. 1998, \apj,
504, L95
                 
\reference{}Burderi, L., Di Salvo, T., D'Antona, F., Robba, N. R.,
Testa, V. 2003, \aap, 404, L43

                 	 
\reference{}Campana, S. \& Stella, L. 2000, \apj, 541, 849

\reference{}Campana, S., Colpi, M., Mereghetti, S., Stella, L.,
Tavani, M. 1998, \aapr, 8, 279
          
\reference{}Campana, S., Stella, L., Gastaldello, F., Mereghetti, S.,
Colpi, M., Israel, G. L., Burderi, L., Di Salvo, T., Robba,
R. N. 2002, \apj, Letters, 575, L15

\reference{}Cash, W. 1979, \apj, 228, 939

\reference{}Chakrabarty, D. \& Morgan, E. H. 1998, \nat, 394, 346

\reference{}Cohn, H. N., Lugger, P. M., Grindlay, J. E., Edmonds, P. D.
2002, \apj, 571, 818

\reference{}Colpi, M., Geppert, U., Page, D., Possenti, A. 2001, \apj, 
548, L175

\reference{}Di Salvo, T. \& Burderi, L. 2003, \aap, 397, 723

\reference{}Edmonds, P. E., Gilliland, R. L., Heinke, C. O., Grindlay, J. E.
2003, \apj, 596, 1197

\reference{}G\"ansicke, B. T., Braje, T. M., \& Romani, R. W. 2002, \aap, 386, 1001

\reference{}Gendre, B., Barret, D., Webb, N. 2003, \aap, 304, L11

\reference{}Grindlay, J. E., Heinke, C., Edmonds, P. D., Murray, S.S. 
2001a, Science, 292, 2290

\reference{}Grindlay, J.~E., Heinke, C.~O., Edmonds,
P.~D., Murray, S.~S., \& Cool, A.~M.\ 2001b, \apjl, 563, L53


\reference{}Heinke, C. O., Edmonds, P. D., Grindlay, J. E., Lloyd, D. A., 
Cohn, H. N., Lugger, P. M. 2003a, \apj, 590, 809

\reference{}Heinke, C. O., Grindlay, J. E., Lugger, P. M., Cohn, H. N., 
Edmonds, P. D., Lloyd, D. A., Cool, A. M., 2003b, ApJ, 598, 501



\reference{}Johnston, H. M., Verbunt, F., \& Hasinger, G. 1995 \aap, 298, 
L21

\reference{}Kong, A. K. H., McClintock, J. E., Garcia, M. R., Murray,
S. S., Barret, D. 2002, \apj, 570, 277

\reference{}Kuulkers, E., den Hartog, P. R., in 't Zand, J. J. M., 
Verbunt, F. W. H. M., Harris, W. E., Cocchi, M. 2003, \aap, 399, 633
 
\reference{}Makishima, K. et al. 1981, \apj, 247, L23

\reference{}Markwardt, C. B. \& Swank, J. H. 2000, \iaucirc,  7454

\reference{}Menou, K. \& McClintock, J. E. 2001, \apj, 557, 304
                 
\reference{}Nowak, M. A., Heinz, S., Begelman, M. C. 2002, \apj,
573, 778


\reference{}Pooley, D.~et al.\ 2002a, \apj, 569, 405
                 	 
\reference{}Pooley, D. et al. 2002b, \apj, 573, 184

\reference{}Rutledge, R. E., Bildsten, L., Brown, E. F., Pavlov, 
G. G., Zavlin, V. E. 2000, \apj, 529, 985

\reference{}Rutledge, R. E., Bildsten, L., Brown, E. F., Pavlov,
G. G., Zavlin, V. E. 2001a, \apj, 551, 921
       
\reference{}Rutledge, R. E., Bildsten, L., Brown, E. F., Pavlov,
G. G., Zavlin, V. E. 2001b, \apj, 559, 1054
                 
\reference{}Rutledge, R. E., Bildsten, L., Brown, E. F., Pavlov,
G. G., Zavlin, V. E., Ushomirsky, G., 2002, \apj, 580, 413
                 
\reference{}Stella, L., Campana, S., Colpi, M., Mereghetti, S.,
Tavani, M., 1994, \apj, 423, L47
                 
\reference{}Stella, L., Campana, S., Merghetti, S., Ricci, D., Israel, G. L.
2000, \apj, 537, L115

\reference{}van Paradijs, J., Verbunt, F., Shafer, R. A., \& Arnaud,
K. A. 1987, \aap, 182, 47
                 
\reference{}Verbunt, F., Bunk, W., Hasinger, G., Johnston, H. M. 
1995, \aap, 300, 732

\reference{}Warwick, R. S., Norton, A. J., Turner, M. J. L., Watson, M. G.,
Willingale, R. 1988, \mnras, 232, 551
      
\reference{}White, N. E. \& Angelini, L. 2001, \apj, 561, L101

\reference{}Wijnands, R. 2003 in ``The Restless 
High-Energy Universe'', 5--8 May 2003, Amsterdam, The Netherlands,
ed. E. P. J. van den Heuvel, J. J. M. in 't Zand, \&
R. A. M. J. Wijers (Elsevier) (astro-ph/0309347; the most up-to-date
version is available at http://zon.wins.uva.nl/$\sim$rudy/admxp/)

\reference{}Wijnands, R., Miller, J. M., Markwardt, C., Lewin,
W. H. G., van der Klis, M. 2001, \apj, 560, L159
       
\reference{}Wijnands, R., Homan, J., \& Remillard, R. 2002, ATEL 101.

\reference{}Wijnands, R., Nowak, M., Miller, J. M., Homan, J., Wachter, S.,
Lewin, W. H. G. 2003, \apj, 594, 952
          
\reference{}
Wijnands, R., Homan, J., Heinke, C. O., Miller, J. M., \& Lewin,
W. H. G., 2004, \apj, submitted


\reference{}Zavlin, V. E., Pavlov, G. G., \& Shibanov, Yu. A., 1996,
\aap, 315, 141
          	
          
\end{references}
\end{document}